\documentclass[Afour,sageh,times]{sagej}

\usepackage{moreverb,url}
\usepackage[colorlinks,bookmarksopen,bookmarksnumbered,citecolor=blue,urlcolor=blue]{hyperref}
\newcommand\BibTeX{{\rmfamily B\kern-.05em \textsc{i\kern-.025em b}\kern-.08em
T\kern-.1667em\lower.7ex\hbox{E}\kern-.125emX}}
\def\volumeyear{2025}
\begin{document}
\title{Free Energy and Network Structure: Breaking Scale-Free Behaviour Through Information Processing Constraints}
\runninghead{Williams and Chen}
\author{Peter R. Williams\affilnum{1,2} and Zhan Chen\affilnum{3}}
\affiliation{\affilnum{1}Rinna KK, Tokyo, Japan\\\affilnum{2}Independent Researcher\\\affilnum{3}Microsoft Japan, Tokyo, Japan}
\email{prw20042004@yahoo.co.uk}

\begin{abstract}
In this paper we show how The Free Energy Principle (FEP) can provide an explanation for why real-world networks deviate from scale-free behaviour, and how these characteristic deviations can emerge from constraints on information processing. We propose a minimal FEP model for node behaviour reveals three distinct regimes: when detection noise dominates, agents seek better information, reducing isolated agents compared to expectations from classical preferential attachment. In the optimal detection regime, super-linear growth emerges from compounded improvements in detection, belief, and action, which produce a preferred cluster scale. Finally, saturation effects occur as limits on the agent's information processing capabilities prevent indefinite cluster growth. These regimes produce the knee-shaped degree distributions observed in real networks, explaining them as signatures of agents with optimal information processing under constraints. We show that agents evolving under FEP principles provides a mechanism for preferential attachment, connecting agent psychology with the macroscopic network features that underpin the structure of real-world networks.
\end{abstract}

\maketitle

\section{Introduction}

Complex networks pervade natural and artificial systems, from cellular interactions to social relationships \citep{Barabasi2016}. A common feature in these networks is the frequent emergence of power-law degree distributions, where the probability of a node having $k$ connections follows $P(k) \propto k^{-\gamma}$ \citep{Newman2003}. The canonical explanation for such distributions is preferential attachment: new nodes preferentially connect to highly connected existing nodes \citep{BarabasiAlbert1999}. While this mechanism successfully describes network growth, it has faced criticism for lacking mechanistic underpinning \citep{Newman2018}. Some researchers argue that power-law distributions might arise from alternative processes \citep{Clauset2009}, suggesting the need for deeper theoretical foundations.

Notably, many real-world networks exhibit systematic deviations from pure power-law behaviour, often showing a characteristic knee shape where moderate-degree nodes are overrepresented relative to both low and high degrees, when compared to the preferential attachment case. While various mechanisms have been proposed to explain these deviations, including aging effects \citep{Dorogovtsev2002} and resource constraints \citep{Amaral2000}, an explanation based on information processing principles has not yet emerged. Understanding these deviations is crucial, as they may reflect constraints on how agents process and act on information in their environment.

Parallel to network science, a different theoretical framework has emerged in cognitive science and neurobiology: the Free Energy Principle (FEP) \citep{Friston2010}. This principle posits that biological systems maintain their order by minimising a quantity called variational free energy, which minimises the surprise (negative log probability) of their sensory states \citep{Friston2013}. FEP provides a mathematical framework for understanding how agents perceive, learn, and act in their environment \citep{Ramstead2018}. Under this principle, biological systems actively sample their environment to confirm their internal models, a process known as active inference \citep{Friston2017}.

Despite their separate development, these two frameworks—-- network formation and free energy minimisation —--might share deeper connections. Both describe systems organising themselves through information-driven processes \citep{Lynn2020, Kirchhoff2018}. While preferential attachment implicitly assumes agents can sense and respond to network structure \citep{Newman2003}, FEP explicitly describes how agents process and act on information \citep{Parr2020}. This suggests that preferential attachment might emerge as a results of more fundamental principles of information processing, an idea supported by recent work on information-theoretic approaches to network formation \citep{Ver2019}.

Recent studies have shown that FEP can explain diverse biological phenomena, from cellular behaviour to neural organisation \citep{Friston2013, Constant2018}. The principle has been successful in bridging scales of organisation \citep{Ramstead2018}, although some researchers question its universal applicability \citep{Colombo2018}. Similarly, while preferential attachment has been observed in many systems \citep{Jeong2003}, from citation networks to protein interactions, its mechanistic origins remain debated \citep{Medo2011}.

A key challenge in both fields is understanding how local, individual-level processes generate global, system-level patterns \citep{Anderson2018}. In network science, this manifests as the emergence of scale-free structures from individual attachment decisions \citep{Bianconi2001}. In FEP, it appears as the generation of adaptive behaviour from local information processing \citep{Kirchhoff2018}. This parallel suggests an opportunity for theoretical synthesis, building on recent work connecting information theory and network formation \citep{Lynn2020}.

Our work bridges these domains by showing that FEP can provide a mechanistic foundation for network formation, revealing how deviations from pure preferential attachment can emerge from information processing constraints. This builds on previous attempts to ground network formation in cognitive processes \citep{Papadopoulos2012} while addressing criticisms of both preferential attachment's lack of mechanism \citep{Newman2018} and FEP's explanatory scope \citep{Colombo2018}. We show that agents following active inference can generate network structures with characteristic deviations from power-law distributions, reflecting regimes of information processing.

The analysis focuses on a simple model where agents move in a one-dimensional space, sensing and responding to their neighbours according to FEP. This model was selected to be sufficiently complex to include all of the key behavioural processes, while staying simple enough to be analytically tractable. Through this model, we identify three distinct information processing regimes - noise-dominated, optimal detection, and saturation - that shape network formation. While similar models have been used to study collective behaviour \citep{Sumpter2010}, our work shows how these regimes can emerge from principles of uncertainty minimisation and create characteristic network structures.

This synthesis has broad implications beyond network science. It suggests that network structures in biological and social systems reflect information processing constraints rather than arbitrary rules or external limitations. The identification of distinct information processing regimes provides new approaches to understanding and designing networks across domains, from social media platforms to networks of artificial agent systems. The framework offers practical insights for network design while deepening our understanding of how cognitive constraints shape collective structure.

In the following sections we develop this theoretical framework. We begin by examining the constraints that biological agents face when processing information about their environment, showing how these create natural scales in network formation. We then derive how agents translate sensory information into movement decisions through free energy minimisation, leading to an attachment kernel that emerges from rational inference under uncertainty. This kernel reveals three distinct regimes of information processing—noise-dominated, optimal detection, and saturation—that shape network structure in characteristic ways. We show how the interplay between these regimes can explain commonly observed deviations from pure scale-free behaviour, particularly the knee-shaped degree distributions found in many real networks. This analysis connects microscopic information processing constraints to the macroscopic network structure, providing a mechanistic foundation for network formation that bridges cognitive and collective behaviour. We conclude by exploring the implications of this framework for understanding diverse systems, from biological networks to artificial agent collectives, and suggest new approaches for analysing and designing networks where information processing plays an important role.

\section{Mathematical Framework Connecting FEP and Preferential Attachment}

We develop a minimal Active Inference model to understand how simple agents learn from and interact with their environment. Our model captures two key processes: how agents update their internal model of the world based on sensory data, and how they choose actions based on this internal model. By using analytically tractable functional forms, we can investigate the emergent properties of collaboration networks that arise from these behavioural patterns.

\subsection{The Agent and its World}

We begin with a simple agent navigating a one-dimensional world through basic left and right movements controlled by velocity commands. The agent's sensory capabilities are limited: it can only detect nearby neighbours with a detection probability that decreases with distance. Internally, the agent maintains a single latent parameter \(b\), which represents its belief about the environmental resource gradient's slope.

The agent model and one-dimensional environment are intentionally minimal to isolate the core principles of information processing constraints on network formation and maintain analytical tractability. This simplification allows us to clearly demonstrate how distinct information processing regimes emerge from FEP principles. We acknowledge that real-world agents and environments are far more complex. Extending the model to higher-dimensional spaces and incorporating richer agent representations would likely modify the quantitative scaling relationships observed. However, we argue that the qualitative regimes – noise-dominated information seeking, optimal detection and super-linear growth, and saturation due to processing limits – and the fundamental mechanism of information processing constraints shaping network structure are likely to be robust features that generalize beyond this simplified setting. Future research could explore these extensions to assess the quantitative impact of increased complexity.

\paragraph{Markov Blanket and States}
The Markov blanket defines the boundary between an agent and its environment. It separates an agent's internal states from external variables, mediating all interaction and information flow between them. Thus the agent's interaction with its environment can be described through two key state types. The agent's \emph{active states} represent motor outputs, that generate velocity \(v_{\text{applied}}\) in the external world, based on the internal latent variable \(b\). This simplifies how an agent's brain states translate into physical movement.

Its \emph{sensory states} consist of binary detection events (``neighbour detected on left'' or ``neighbour detected on right''), providing noisy information about neighbour positions. Beyond these states lie what we call the external variables, including the agent's true position and the actual resource distribution. These remain unknowable to the agent, existing outside its Markov blanket.

\paragraph{Choice of Prior}
The agent's prior distribution \(\pi(b)\) on the slope parameter \(b\) represents its baseline expectations about typical resource gradient behaviour. In biological terms, this prior might be genetically encoded through evolutionary processes, developed during early learning phases, and remain fixed throughout the agent's lifetime (for model simplicity). While our current model assumes a static prior, future extensions could incorporate prior adaptation based on accumulated experience.

\paragraph{The Likelihood Function}
The likelihood function \(L(\text{Obs}\mid b)\) serves as the bridge between external reality and internal model, mapping how sensory readings arise from hypothesised environmental states. Much like how an eye transforms incoming photons into neural activity, this function translates external states (neighbour detection) into the agent's internal model (beliefs about resource gradients).

In our specific implementation, the agent processes neighbour-detection events by counting detections over a brief time window and compressing these counts into a scalar statistic \(D\) (e.g., right-side detections minus left-side detections). We use \(D\) as the primary sensory reading, where a large positive value suggests neighbour clustering in the positive direction, showing a larger \(b\) through our likelihood function definition.

\paragraph{A Continuously Updating Variational Posterior}
The recognition density, or variational posterior \(Q(b)\), represents the agent's current beliefs about the environmental gradient after observing sensory data. This probability distribution over possible slope values \(b\) continuously updates as new information arrives through the sensory apparatus. The recognition density serves as a computationally tractable approximation to the true Bayesian posterior, allowing the agent to maintain and update beliefs in real-time without requiring exhaustive computation of the complete posterior distribution.

\paragraph{The Free Energy and Minimisation}
The free energy \(F\) originates from the principle of least action in statistical physics, generalised to biological systems. It represents the difference between the agent's internal model of reality and the actual environmental dynamics. Through variational calculus, we can derive the specific form relevant to our model,
\begin{equation}
  F\bigl[Q(b)\bigr] = \int Q(b)\ln \frac{Q(b)}{\pi(b)}db - \int Q(b)\ln L\bigl(\text{Obs}\mid b\bigr)db.
  \label{eq:free_energy_final}
\end{equation}
The first term, known as the Kullback-Leibler divergence, measures how far the current beliefs deviate from prior expectations. The second term evaluates how well these beliefs explain incoming sensory data.

Behaviourally, this creates a trade-off between maintaining consistency with evolutionary and developmental knowledge (first term) and adapting beliefs about the environmental gradient based on new sensory information (second term). The agent updates its internal model by minimising \(F\), which involves adjusting \(Q(b)\) to balance prior knowledge with new sensory evidence. When sensory data is uninformative, the posterior stays close to the prior; when strong evidence arrives, the posterior adapts to better explain the new observations.

\paragraph{Action and External Movement}
The agent's internal belief about the environmental gradient, represented by the parameter \(b\), guides its physical movement through the environment. This translation from internal state to external action occurs through what we might think of as the agent's motor system. In this model, the action-generation process of the agent involves generating a velocity based on its current best estimate of the environmental gradient,
\begin{equation}
  v = f\bigl(\mu_b\bigr),
\end{equation}
where \(f\) is a mapping function that converts the posterior mean belief \(\mu_b\) into a desired movement speed and direction. This function represents the agent's behavioural policy, i.e. how it responds to its beliefs about the environment. For example, a simple linear mapping would cause the agent to move faster in environments where it believes the resource gradient is steeper,
\begin{equation}
v = \gamma\mu_b.
\end{equation}
Further refinements may involve factoring in that perfect execution of intended movements is impossible in biological and non-biological systems. Motor noise, environmental perturbations, and imperfect muscle control could all be included here in a more complex model.

This action-generation process completes the active inference loop: the agent's internal beliefs about the gradient \(\mu_b\) drive its physical movement \(v\), which in turn changes its position in the environment. This repositioning alters the spatial relationships with neighbours, leading to new patterns of detection events that update the agent's beliefs through the free energy minimisation process described earlier. Through this continuous cycle of perception and action, the agent maintains an ongoing dynamic relationship with its environment, constantly updating its model and adjusting its behaviour in response to new evidence.

\subsection{Closed-Form Expression of \texorpdfstring{$\mu_b$}{μb} under Gaussian Assumptions}

To derive tractable expressions for the free energy, we make several assumptions about the functional forms of our distributions. While these assumptions primarily serve mathematical convenience, they maintain plausibility with biological agents while enabling analytical solutions to the free energy minimisation problem.

We begin with the prior distribution \(\pi(b)\), which we model as a Gaussian distribution
\begin{equation}
  \pi(b) = \mathcal{N}\bigl(b;\mu_\pi,\sigma_\pi^2\bigr).
\end{equation}
The parameters of this distribution encode the agent's baseline expectations about environmental gradients. The mean \(\mu_\pi\) represents the expected slope magnitude, where \(\mu_\pi = 0\) shows no prior belief in a gradient's existence, while \(\mu_\pi > 0\) suggests an innate expectation of positive gradients. The variance \(\sigma_\pi^2\) quantifies the flexibility of these prior beliefs. While the Gaussian form is not mandated by the Free Energy Principle, it aligns with standard approaches in variational Bayes and provides analytically tractable solutions.

For the likelihood function, we approximate the distribution of the detection statistic \(D\) using a Gaussian whose mean depends linearly on the slope parameter \(b\),
\begin{equation}
  L\bigl(\text{Obs}\mid b\bigr) = \mathcal{N}\!\Bigl(D;\mu_{D}(b),\sigma_{D}^2\Bigr),
\end{equation}
where \(\mu_{D}(b) = \alpha\,b + \beta\). This linear relationship captures how larger environmental gradients lead to higher expected detection counts. The coefficient \(\alpha\) determines the sensitivity of the detection statistic to changes in the slope, while \(\beta\) accounts for any baseline bias in the detection process. The variance \(\sigma_{D}^2\) represents various sources of uncertainty in the sensory process, including imperfect detection abilities, environmental noise, and finite sampling windows. While more complex functional forms could model the relationship between slope and detections, this linear Gaussian approximation provides a reasonable balance between biological realism and mathematical tractability.

Finally, we model the recognition density \(Q(b)\) as a Gaussian distribution
\begin{equation}
  Q(b) = \mathcal{N}\bigl(b;\mu_b,\sigma_b^2\bigr).
\end{equation}
This distribution represents the agent's current belief state about the environmental gradient. The mean \(\mu_b\) encodes the agent's best estimate of the slope. A large positive \(\mu_b\) indicates a belief in resources increasing in the positive \(x\)-direction. The variance \(\sigma_b^2\) quantifies the agent's uncertainty about this estimate, with smaller values indicating greater confidence. This Gaussian form for \(Q(b)\) enables the agent to maintain a computationally efficient representation of its beliefs using just two parameters, \(\mu_b\) and \(\sigma_b^2\), which update as new sensory information arrives.

The combination of these Gaussian assumptions leads to closed-form expressions for both terms in the free energy functional. This mathematical convenience allows us to derive explicit update equations for the recognition density parameters, making the model both analytically tractable and biologically interpretable. While alternative distributional choices could offer more precise models of biological reality, our Gaussian assumptions capture the essential features of belief updating while maintaining mathematical simplicity.

Under these Gaussian assumptions for the prior, likelihood, and recognition density, the free energy functional admits an explicit closed form (see the Appendix at the end of this paper for the complete derivation). The gradient of the free energy with respect to the agent's current estimate of the slope takes the form
\begin{equation}
\label{eq:dF_db}
\frac{\partial F}{\partial \mu_b} = \frac{\mu_b - \mu_\pi}{\sigma_\pi^2} + \frac{-\,\alpha\,(D-\beta) + \alpha^2\,\mu_b}{\sigma_D^2},
\end{equation}
where $\mu_b$ represents the agent's current slope estimate, $\mu_\pi$ its prior expectation, and $D$ the observed detection statistic. Setting this gradient to zero yields the solution for the optimal slope estimate
\begin{equation}
\mu_b = \frac{(\alpha/\sigma_D^2)\,(D-\beta) \;+\; \mu_\pi/\sigma_\pi^2}{\alpha^2/\sigma_D^2 + 1/\sigma_\pi^2}.
\end{equation}
This expression reveals how the agent balances prior knowledge against sensory evidence when updating its beliefs. The denominator terms $\alpha^2/\sigma_D^2$ and $1/\sigma_\pi^2$ act as precision weights: when sensory noise $\sigma_D^2$ is large, the agent relies more on its prior beliefs. Conversely, when prior uncertainty $\sigma_\pi^2$ is large, it weights new evidence more. The parameter $\alpha$ determines how sensitive the detection statistic is to the environmental gradient: a larger $\alpha$ means that changes in the slope produce larger changes in neighbour detection patterns, amplifying the influence of sensory data on belief updates. This mathematical structure implements a form of optimal Bayesian updating that could plausibly be approximated by neural circuits in biological systems, but also provides a concrete example of how free energy minimisation might be realised in a model agent.

While the assumption of Gaussian distributions for the prior, likelihood, and recognition density is primarily motivated by analytical tractability, it also possesses some plausibility. Aggregated sensory data, as considered in our detection statistic, may indeed approach Gaussian distributions due to the central limit theorem. Furthermore, Gaussian distributions provide a parsimonious and mathematically convenient representation of uncertainty, which aligns with variational Bayesian methods. However, we acknowledge that real-world noise and belief distributions may deviate from perfect Gaussianity, and future work could explore the impact of non-Gaussian noise models on the quantitative aspects of network formation.

\section{Deriving the Attachment Kernel and Information Processing Regimes}

The emergence of network structure in our model can be understood through how agents detect and respond to clusters of neighbours. To develop this understanding, we first analyse the detection process itself, then examine how detection information shapes movement decisions, and finally identify distinct regimes of information processing that characterise network formation.

\subsection{From Detection to Movement}

Consider first how an agent detects a cluster of $d$ nearby agents. Each cluster member has a probability $p$ of being detected within the agent's sensing window $\tau$. These detection events can be modeled as independent Bernoulli trials, leading to a binomial distribution of detection counts. For clusters above some critical size, the central limit theorem implies these counts approach a Gaussian distribution, giving our detection statistic
\begin{equation}
D \sim \mathcal{N}(\alpha d,\eta^2)
\end{equation}
where $\alpha = p/\tau$ represents the base detection rate per cluster member and $\eta^2$ captures both intrinsic sensing noise and temporal fluctuations in detection counts.

This detection signal feeds into the agent's belief updating process through our free energy minimisation framework. For large clusters where the detection signal dominates, $\alpha d \gg \eta$, the posterior mean belief about environmental gradients approaches
\begin{equation}
\mu_b \approx \frac{(\alpha^2/\sigma_D^2)d}{\alpha^2/\sigma_D^2 + 1/\sigma_\pi^2} \approx C d
\end{equation}
where
\begin{equation}
C = \frac{\alpha^2\sigma_\pi^2}{\alpha^2\sigma_\pi^2 + \sigma_D^2}
\end{equation}
represents the coupling strength between cluster size and belief formation.

These beliefs about environmental gradients drive movement through the agent's velocity control,
\begin{equation}
v = \gamma\mu_b \propto d.
\end{equation}
This creates a relationship: larger clusters generate stronger detection signals, leading to stronger gradient beliefs and faster directed movement.

\subsection{Constraints in Biological Agents}

Before examining how agents form networks through their interactions, we must first understand the inherent limitations that any biological agent faces when processing and acting on information. These constraints create natural scales in network formation, shaping the degree distribution in ways that deviate from pure preferential attachment.

Our agent faces three limitations. First, it cannot maintain arbitrarily strong beliefs about its environment. The Gaussian prior on environmental gradients $\pi(b)=\mathcal{N}(b;\mu_\pi,\sigma_\pi^2)$ imposes this constraint: beliefs about gradients much larger than $\sigma_\pi$ incur heavy penalties through the KL-divergence term in the Free Energy formulation. This creates an effective upper bound $b_{\text{max}}$ on believable gradient values.

Second, the agent has finite sensory capabilities. No agent can detect an unlimited number of neighbours or process an infinite stream of sensory information. Our agent model captures this through finite detection ranges, which impose a maximum $k_{\text{max}}$ on the number of neighbours that can be simultaneously detected in any direction.

Third, the agent cannot move at arbitrary speeds, even if it detects a strong environmental gradient. The relationship between inference and action in our model $v = \gamma\mu_b$ combines with the bound on believable gradients to create a maximum achievable velocity $v_{\text{max}} = \gamma b_{\text{max}}$.

These limitations manifest as characteristic scales in the degree distribution. From the bound on believable gradients, we obtain $d_{\text{belief}} \approx b_{\text{max}}/C$, beyond which additional cluster members cannot produce proportionally stronger beliefs. The sensory limitation creates a second scale $d_{\text{sensory}} = k_{\text{max}}$ where detection saturates. The velocity bound gives us a third scale $d_{\text{ability}} = v_{\text{max}}/\gamma C$ beyond which movement speed cannot increase further.

\subsection{Deriving the Attachment Kernel}

To understand how network structure emerges from individual behaviour, we must connect an agent's information processing capabilities to its probability of joining clusters of different sizes. This connection, which we call the attachment kernel, provides the mechanistic link between microscopic behaviour and macroscopic network structure.

\paragraph{From Detection to Belief}
Consider an agent encountering a cluster of $d$ agents. Each cluster member generates detection events with probability $p$ within the agent's sensing window $\tau$. These independent detection events follow a binomial distribution which, for clusters above a critical size ($d \gtrsim 10$), approaches a Gaussian by the central limit theorem:
\begin{equation}
D \sim \mathcal{N}(\alpha d, \eta^2)
\end{equation}
where $\alpha = p/\tau$ represents the base detection rate and $\eta^2$ captures both intrinsic sensing noise and temporal fluctuations in the detection counts.

This detection statistic shapes the agent's beliefs through free energy minimisation. For large clusters where the detection signal dominates ($\alpha d \gg \eta$), we can substitute our expression for $D$ into Equation \ref{eq:dF_db} to find the posterior mean belief about environmental gradients:
\begin{equation}
\mu_b \approx \frac{(\alpha^2/\sigma_D^2)d}{\alpha^2/\sigma_D^2 + 1/\sigma_\pi^2} \approx C d
\end{equation}
where
\begin{equation}
C = \frac{\alpha^2\sigma_\pi^2}{\alpha^2\sigma_\pi^2 + \sigma_D^2}
\end{equation}
represents the coupling between cluster size and belief strength. This linear relationship between $\mu_b$ and $d$ emerges from: (i) the physics of neighbour detection, (ii) our Gaussian modelling assumptions, and (iii) the dominance of detection signals for large clusters.

\paragraph{From Belief to Movement}
The agent's belief about environmental gradients drives movement through its velocity control,
\begin{equation}
  v_{\text{intended}} = \gamma\mu_b \propto d
\end{equation}
where $\gamma$ controls movement responsiveness. This creates a behavioural pattern: agents move more quickly toward larger clusters because stronger detection signals generate stronger beliefs about resource gradients in that direction.

\paragraph{From Movement to Attachment}
To connect movement with attachment probability, consider the time $T_{\text{move}}(d)$ required for an agent to reach a cluster. This time depends inversely on velocity,
\begin{equation}
T_{\text{move}}(d) \approx \frac{l}{v} \propto \frac{1}{d}
\end{equation}
where $l$ represents a characteristic distance. The attachment kernel $P(\text{attach}\mid d)$, the probability that an agent successfully joins a cluster, should scale inversely with this convergence time,
\begin{equation}
P(\text{attach}\mid d) \propto \frac{1}{T_{\text{move}}(d)} \propto d.
\end{equation}

While our framework shares similarities with preferential attachment in describing a ``rich-get-richer'' dynamic, it is crucial to emphasize that it provides a mechanistic foundation rooted in agent behavior and information processing, rather than positing preferential attachment as an axiomatic rule. In our model, preferential attachment, particularly the linear scaling regime, emerges as a consequence of agents minimizing free energy under optimal detection conditions. However, the key contribution of our framework lies in explaining deviations from pure scale-free behavior. The information processing constraints and the resulting noise-dominated and saturation regimes are what differentiate our approach from traditional preferential attachment models, explaining the empirically observed knee-shaped degree distributions as a modification and refinement, rather than a complete break, from the scale-free archetype.

\paragraph{Detection Time and Super-linearity}
The total time to attach includes both detection and movement phases. For a cluster of size $d$, the detection time scales as $T_{\text{det}}(d) \sim d^{-\beta/2}$ since signal-to-noise ratio improves as $\sqrt{d}$, while movement time scales as $T_{\text{move}}(d) \sim 1/d$. The total attachment time $T(d) = T_{\text{det}}(d) + T_{\text{move}}(d) \sim d^{-\beta/2} + 1/d$ can thus decrease faster than $1/d$ when $\beta \geq 2$, leading to attachment probabilities $P(\text{attach}\mid d) \sim 1/T(d)$ that grow super-linearly with $d$. This super-linear scaling emerges from the compound effect of improved detection efficiency and faster movement toward larger clusters, operating most effectively in the intermediate regime between noise-dominated detection and saturation.

This derivation reveals how in the absence of constraints on the agent's abilities, linear preferential attachment emerges from biological information processing rather than being imposed as an external rule. When agents follow free energy minimisation principles, they generate the ``rich get richer'' phenomenon: larger clusters create stronger detection signals, leading to stronger gradient beliefs and faster directed movement, producing attachment probabilities that scale linearly with cluster size.

This base linear scaling will be modified by the constraints we identified earlier and the information processing regimes we discuss next. However, understanding this core mechanism, from detection through belief to movement and finally attachment, provides the foundation for analysing how network structure emerges from individual behaviour.

\subsection{Characterising Information Processing Regimes}

The FEP agent model identifies three characteristic scales arising from limitations:
\begin{align}
d_{\text{belief}} &\approx b_{\text{max}}/C \quad &\text{(prior belief constraint)} \\
d_{\text{sensory}} &= k_{\text{max}} \quad &\text{(sensory limitation)} \\
d_{\text{ability}} &= v_{\text{max}}/\gamma C \quad &\text{(velocity bound)}.
\end{align}
Using results from the previous section, these characteristic scales can be expressed in terms of the FEP model parameters,
\begin{align}
d_{\text{belief}} &= \frac{b_{\text{max}}}{C} = \frac{\sigma_\pi}{\alpha}\left(1 + \frac{\sigma_D^2}{\alpha^2\sigma_\pi^2}\right) \\
d_{\text{sensory}} &= k_{\text{max}}, \\
d_{\text{ability}} &= \frac{v_{\text{max}}}{\gamma C} = \frac{v_{\text{max}}}{\gamma}\left(1 + \frac{\sigma_D^2}{\alpha^2\sigma_\pi^2}\right),
\end{align}
showing how these scales emerge from the interplay between detection efficiency ($\alpha$), measurement noise ($\sigma_D$), prior uncertainty ($\sigma_\pi$), and movement responsiveness ($\gamma$). These limitations define three distinct regimes:

\paragraph{Noise-Dominated:} When $d \lesssim \eta/\alpha$, weak detection signals lead to high posterior uncertainty $\sigma^2_b$. This uncertainty drives exploratory behaviour through belief-dependent movement noise $\eta_v \sim \mathcal{N}(0, \sigma^2_b)$, causing agents to seek better information to reduce their uncertainty.

\paragraph{Optimal Detection:} For intermediate cluster sizes, the three mechanisms
\begin{align}
L(\text{Obs} \mid b) =& \mathcal{N}(\alpha d + \beta, \sigma^2_D) \nonumber\\
                     & \text{(linear detection scaling)}, \\
                     & \nonumber\\
\sigma^2_b =& \left(\frac{1}{\sigma^2_\pi} + \frac{\alpha^2}{\sigma^2_D}\right)^{-1} \nonumber\\
                     & \text{(improved precision), and} \\
                     & \nonumber\\
v =& \gamma\mu_b + \eta_v, \quad \eta_v \sim \mathcal{N}(0, \sigma^2_b) \nonumber\\
                     & \text{(accurate movement)}
\end{align}
compound. This creates a positive feedback loop, producing super-linear growth in attachment probability.

\paragraph{Saturation:} When $d$ exceeds any of the limits $(d_{\text{belief}}, d_{\text{sensory}}, d_{\text{ability}})$, growth slows due to: sensory saturation; inability to detect additional neighbours beyond $k_{\text{max}}$; prior belief constraints; implausible gradients beyond $b_{\text{max}}$; and movement limitations; or velocities cannot exceed $v_{\text{max}}$.

\paragraph{Degree Distribution Effects:} These regimes create natural preferred scales in the degree distribution. Information-seeking in the noise-dominated regime reduces the number of low-degree nodes compared to preferential attachment. At high degrees, the three limitations prevent indefinite growth, reducing the number of high-degree nodes. These preferred scales emerge from the information processing constraints inherent in the FEP framework.

\subsection{Network Structure Through the Lens of Information Processing}

The degree distribution that emerges from our free energy minimising agents can be understood through how they can process and act on information about their environment. Three distinct information processing regimes create the characteristic network structure we observe.

\paragraph{Information Sparsity}
In environments with small clusters, agents face an information processing challenge. Their detection statistics are dominated by noise, making it difficult to form reliable beliefs about gradient direction. In this regime, where $\alpha d \lesssim \eta$, agents must rely on their prior beliefs to guide behaviour. These prior beliefs might encourage more connections than pure preferential attachment would predict, as agents seek to improve their ability to sense their environment through increased connectivity. This explains why we observe fewer isolated nodes than the Barabási-Albert model would suggest agents in information-poor environments are driven to increase their degree beyond what random chance would dictate.

\paragraph{Information Abundance}
As cluster sizes increase to moderate levels, agents enter an information processing ``sweet spot.'' Here, the signal-to-noise ratio becomes favourable, but clusters remain small enough that all sensory and cognitive mechanisms can operate effectively. In this regime, agents can form strong, reliable beliefs about environmental gradients and generate appropriate movement responses. The enhanced certainty in this regime makes such clusters attractive targets for attachment. This creates a concentration of nodes around these optimal cluster sizes, forming the characteristic knee in the degree distribution. This knee represents not just a mathematical feature but a biological optimum where agents' information processing capabilities are best matched to their environment.

\paragraph{Information Saturation}
Finally, when clusters become very large, agents enter a regime where they cannot process all available information. Whether limited by sensory capacity \(k_{\text{max}}\), prior beliefs about plausible gradients \(b_{\text{max}}\), or movement capabilities \(v_{\text{max}}\), agents become unable to respond proportionally to further increases in cluster size. In this regime, the attachment probability grows more slowly than cluster size, creating fewer highly connected nodes than pure preferential attachment would predict. This saturation reflects limitations in biological information processing rather than a lack of information.

\paragraph{Emergence of Network Structure with Preferred Scales}
These three information processing regimes: sparse, optimal, and saturated, shape the network's degree distribution. The resulting structure shows fewer isolated nodes than pure preferential attachment (as agents seek better information), a concentration of nodes around optimal cluster sizes (where information processing works best), and fewer high-degree nodes (where information processing saturates). This pattern emerges not from imposed rules but from principles of how agents process and act on environmental information through free energy minimisation.

\paragraph{Scale-Free Networks as Evidence of Different Information Processing Dynamics}
The existence of true scale-free networks provides an illuminating contrast to our framework. When a network exhibits a pure power-law degree distribution, it tells us something fundamental about its nodes: they must operate outside the information processing constraints ($k_{\text{max}}$, $b_{\text{max}}$, $v_{\text{max}}$) that characterise biological agents. These nodes somehow overcome the limitations that create saturation effects in large clusters.

This observation suggests two mechanisms for the emergence of scale-free structure. First, the nodes might possess information processing capabilities that scale with cluster size, pushing $d_*$ to arbitrarily large values. Alternatively, the network's growth might be driven by factors independent of local information processing—for instance, when attachment decisions rely on global rather than local information, bypassing the constraints of direct neighbour detection and response.

Our framework thus reframes the distinction between scale-free and biological networks in terms of information processing. While biological networks emerge from agents working within fixed information processing constraints, scale-free networks reflect dynamics that transcend these limitations. This perspective suggests a testable prediction: if nodes in a biological network could enhance their information processing capabilities—increasing $k_{\text{max}}$, $b_{\text{max}}$, and $v_{\text{max}}$—the resulting degree distribution should become more scale-free as saturation effects diminish.

\subsection{The Driving Power Law behaviour Underpinning the Degree Distribution}

The FEP framework predicts distinct attachment dynamics in each information processing regime that drives the formation of network structure. While analytical derivation of the complete degree distribution is intractable due to the complex interplay between regimes, we can analyse the driving attachment behaviour within each regime to understand how they combine to shape network evolution. While these driving behaviours do not manifest as observable degree distributions, they shape how the network evolves through their combined effects.

\paragraph{Noise-Dominated Regime:}
When detection noise dominates ($\alpha d \lesssim \eta$), agents cannot reliably sense cluster size, leading to approximately random attachment behaviour
\begin{equation}
P(\text{attach}|k) \sim \text{const.}
\end{equation}
This flat scaling reflects agents exploring their environment to gather better information rather than exhibiting strong size preferences.

\paragraph{Optimal Detection Regime:}
In the intermediate regime where information processing operates effectively, three mechanisms compound to produce super-linear attachment
\begin{equation}
P(\text{attach}|k) \sim k^\nu, \quad \nu > 1
\end{equation}
This super-linear scaling emerges from the linear increase in detection signal strength with cluster size, improved belief precision as more evidence accumulates, and more accurate movement toward larger targets. The super-linearity reflects how better detection enables more precise movement, which facilitates more reliable detection - a positive feedback loop in information processing efficiency.

\paragraph{Saturated Regime:}
As cluster size approaches the limitations ($k_{\text{max}}$, $b_{\text{max}}$, $v_{\text{max}}$), information processing saturates at $k_\star = \min(d_{\text{belief}}, d_{\text{sensory}}, d_{\text{ability}})$. From maximum entropy principles, when a system has constrained resources, we expect an exponential-like cutoff in attachment probability
\begin{equation}
P(\text{attach}|k) \sim e^{-k} \quad \text{for } k \gtrsim k_\star.
\end{equation}

\paragraph{Emergent Degree Distribution}
These three regimes of attachment dynamics combine as agents encounter clusters of different sizes. The observed degree distribution reflects this continuous transition between regimes: from random attachment at low degrees ($k^0$), through a congested intermediate range where super-linear growth ($k^\nu$, $\nu>1$) concentrates nodes that would otherwise remain at lower degrees, to exponential suppression $e^{-k}$ at high degrees. This progression produces the characteristic knee shape in the degree distribution, deviating from pure scale-free behaviour in ways that reveal the constraints on information processing through the Free Energy Principle.

\section{Discussion}

The analysis above provides a theoretical framework that connects the Free Energy Principle (FEP) \citep{Friston2010} to the emergence of network structures through a process different from pure preferential attachment. While preferential attachment has long been known to generate power-law distributions in many systems \citep{BarabasiAlbert1999, Newman2003}, our work shows how biological information processing constraints can lead to deviations from scale-free behaviour. By demonstrating that network structures emerge from agents minimising variational free energy, we offer an explanation that links individual cognitive and perceptual processes to large-scale network topologies. This synthesis addresses longstanding criticisms regarding the lack of mechanistic foundations in network formation models \citep{Clauset2009}, and bridges the gap between psychological or cognitive principles and collective structural patterns \citep{Anderson2018}.

Our derivation reveals how the gradients of free energy drive agents toward information-rich regions through three distinct regimes of information processing. In the noise-dominated regime, agents seek better information, leading to fewer isolated nodes than traditional preferential attachment would predict. In the optimal detection regime, a super-linear growth mechanism emerges from the cascade of improving detection statistics, belief precision, and movement accuracy. This super-linearity, arising from free energy minimisation rather than imposed rules, creates a characteristic concentration of nodes around optimal cluster sizes. Finally, in the saturation regime, limitations in agents' ability to process information prevent indefinite cluster growth, leading to fewer high-degree nodes than pure preferential attachment would predict.

These findings align with empirical observations that real networks often exhibit finite-size effects, exponential cutoffs, and deviations from exact power laws \citep{Broido2019}. Our framework shows these deviations are not imperfections but signatures of how agents process and act on environmental information. The characteristic ``knee-shaped'' degree distributions, where moderate-degree nodes are overrepresented relative to both low and high degrees, emerge from the interplay between information seeking in the sparse regime, super-linear growth in the optimal regime, and saturation at large scales.

Previous work has noted that information-based constraints may drive deviations from scale-free structure, particularly in biological and financial networks. In neural systems, network topology appears to reflect a trade-off between information transmission efficiency and the metabolic costs of maintaining connections \citep{lynn2020physics, avena2018communication}. Similar patterns emerge in financial networks, where information processing capacity constraints lead to characteristic deviations from pure power laws \citep{bardoscia2021physics}. Studies of brain dynamics suggest that local information processing limitations can have substantial effects on global network structure \citep{gollo2018diversity}. Our framework provides a novel perspective by showing how these deviations can emerge from the process of free energy minimisation, with no need to model communication channels or metabolic constraints. The three information processing regimes we identify—-- noise-dominated, optimal detection, and saturation—-- provide a mechanistic explanation for how individual cognitive constraints shape collective network structure, unifying observations across multiple domains through the lens of uncertainty minimisation.

Compared to existing explanations, many prior models rely on external constraints or additional parameters. For instance, fitness-based models endow nodes with intrinsic attributes that modify attachment probabilities \citep{Bianconi2001}, while aging models reduce the attractiveness of older nodes \citep{Dorogovtsev2002}, and resource-limited models impose exogenous capacities to induce truncation \citep{Amaral2000, Newman2005}. Although effective at reproducing empirical distributions, these approaches often lack a direct link to the internal, information-processing motivations of agents. Uur FEP-based derivation places the origin of non-scale-free degree distributions in principles of uncertainty minimisation, inference, and action selection.

This perspective helps reinterpret network structures not simply as structural curiosities, but as emergent consequences of cognition and perception operating under uncertainty constraints. It aligns with the notion that living systems exploit information for adaptive behaviour, connecting network organisation to questions about the relationship between information and organisation in biological systems \citep{England2015}. The identification of distinct information processing regimes suggests network structure may serve as a signature of underlying cognitive architectures, resonating with work suggesting that agents evolve to optimise their information processing capabilities \citep{Friston2013, Constant2018}.

Our framework provides concrete, testable hypotheses about how sensory noise $\eta$, detection efficiency $\alpha$, and biological constraints $d_\star$ should influence network formation. Empirical work could measure how changes in these parameters affect network growth, potentially testing these predictions in controlled biological or social experiments. This offers a path toward bridging theory and empiricism, moving beyond phenomenological fits to identify underlying principles that govern the formation of complex networks. It also suggests that carefully manipulating information processing parameters could ``engineer'' desired network structures in artificial systems, such as swarm robotics \citep{Hamann2018} or distributed computing \citep{Jelasity2007}, particularly in scenarios where optimal information processing is crucial.

\subsection{Model Assumptions and Limitations}

Our theoretical framework, while capturing key features of network formation through free energy minimisation, relies on several key assumptions that warrant careful examination. The choice of Gaussian distributions for the prior $\pi(b)$, likelihood $L(\text{Obs} \mid b)$, and recognition density $Q(b)$ enables analytical tractability but represents an idealisation of real biological systems. While the central limit theorem suggests that aggregated detection events may indeed approach Gaussian distributions for large clusters, the true distributions, particularly in sparse networks or at small scales, may exhibit significant deviations from normality. However, there is a robustness to the generalisable quality to the constraints we identify: information sparsity at low degrees, optimal processing at intermediate scales, and saturation at high degrees emerge from basic principles of uncertainty minimisation rather than specific distributional choices. While Gaussian assumptions enable analytical tractability, it is important to acknowledge that deviations from Gaussianity in real systems could influence quantitative predictions, an area for future investigation.

The one-dimensional spatial model we employ represents another important simplification of real-world network formation dynamics. While higher-dimensional spaces would more accurately reflect the embedding of most natural and social networks, the one-dimensional case captures the essential features of gradient-following behaviour under uncertainty. The key mechanisms we identify: the relationship between cluster size and detection statistics $D \sim \alpha d + \eta$, the emergence of super-linear growth in the optimal detection regime, and the existence of natural saturation scales $d_\star$, generalise to higher dimensions, though with modified scaling relationships. Indeed, the dimensional reduction inherent in many real networks, where agents often respond to scalar measures of connection quality or social distance \cite{papadopoulos2012popularity}, suggests that our one-dimensional analysis may capture more of the relevant physics than might at first appear. Future work could explore how the addition of spatial dimensions modifies the quantitative predictions while preserving the qualitative structure of our results.

\subsection{Universal Features of Free Energy Networks}

The three information processing regimes identified here, noise-dominated, optimal detection, and saturation, can emerge from the structure of free energy minimisation rather than from the specific forms of our recognition density $Q$ or likelihood function $L$. This universality suggests that similar behavioural patterns should appear whenever agents process information to guide actions under uncertainty, regardless of the particular domain. While the mathematical details may vary, the core dynamics of seeking better information in sparse environments, experiencing super-linear benefits in optimal regimes, and hitting processing limits in dense regions appear to be general features of free energy minimising systems.

This perspective offers new insights across diverse fields where networked behaviour emerges from individual agents processing and acting on information. In evolutionary biology, the formation and size distribution of animal groups has long puzzled researchers \citep{Couzin2009}. Our framework suggests that observed patterns - from bacterial colonies to fish schools to primate groups - may reflect information processing constraints rather than just environmental pressures. The tendency of many species to maintain specific group sizes might represent an evolved optimisation for the ``sweet spot'' regime, where social information processing is most effective \citep{Sumpter2010}. Similarly, the emergence of hierarchical structures in many animal societies could reflect natural solutions to information processing constraints as groups grow beyond the optimal detection regime \citep{Flack2012}.

In economic networks, our framework offers a novel perspective on classical questions of industrial organisation. Coase's \citep{Coase1937} insight about transaction costs determining firm boundaries might be reinterpreted through the lens of information processing regimes. The observed size distribution of firms, with its characteristic deviations from power laws \citep{Axtell2001}, could reflect the interplay between improved coordination in the optimal regime and degraded information processing in the saturation regime. This view also suggests new approaches to understanding market structure evolution \citep{Jackson2010} and the emergence of supply chain networks \citep{Schweitzer2009}, where information processing constraints may shape network topology as much as traditional economic factors.

Urban systems provide another domain where information processing constraints appear to shape network formation. The widely observed scaling laws in city growth and structure \citep{Bettencourt2013} might reflect transitions between information processing regimes as cities grow. The emergence of polycentric urban forms and the limits to efficient city size could be understood through our framework's prediction of saturation effects in large networks. Transportation network design, too, might benefit from considering how information processing constraints affect human movement and interaction patterns \citep{Batty2013}.

In ecological networks, the framework offers insights into species interaction patterns and ecosystem stability. The observed structure of food webs and mutualistic networks, which often deviate from pure scale-free topologies \citep{Bascompte2009}, might reflect constraints on how agents can process and respond to information about resource availability and potential interactions. This perspective could help explain the emergence of modularity in ecological networks \citep{Olesen2007} and provide new approaches to understanding ecosystem resilience.

These applications highlight how the Free Energy Principle can provide mechanistic explanations for widely observed network phenomena. By focusing on the constraints of information processing rather than domain-specific mechanisms, our framework offers a unifying perspective on network formation across scales and contexts. This suggests that future research might benefit from considering how information processing constraints shape network evolution, potentially leading to more effective interventions in these various systems.

\subsection{Empirical Support from Existing Studies}

Our theoretical framework makes several key predictions about network formation that find substantial support in empirical studies across multiple domains. The predicted knee-shaped degree distribution has been well documented in both biological and social networks. For instance, \cite{kossinets2006empirical} analysed a large-scale email communication network ($n > 40,000$), finding degree distributions that deviate from power-law behaviour in ways that align with our predicted information processing regimes. This knee-shape, with an overrepresentation of moderate-degree nodes, directly aligns with our predicted outcome of the optimal detection regime, where agents concentrate connections around cluster sizes that best match their information processing capabilities. Similar patterns emerge in protein interaction networks \cite{jeong2003lethality}, where degree distributions show clear saturation effects at high degrees. 

The transitions between our predicted regimes find particular support in studies of collective animal behaviour. \cite{sumpter2010collective}'s analysis reveals distinct phases in group formation that parallel our theoretical predictions: an initial noise-dominated phase where individuals seek group membership, an optimal detection phase characterised by rapid growth, and a saturation phase where group size stabilises. Notably, many species maintain group sizes within specific ranges that may correspond to our predicted optimal detection regime (\(\alpha d \gg \eta\) but \(d \ll d_\star\)), where information processing effectiveness peaks.

Our framework provides mechanistic explanations for well-documented social phenomena. Dunbar's number \citep{Dunbar1992, dunbar2016human}, the upper limit on stable social relationships (~150 individuals), aligns with our saturation regime predictions where information processing constraints limit network growth. Beyond a certain network size, information processing constraints, such as limited cognitive capacity, prevent further proportional growth in network connectivity, resulting in a plateau and deviation from scale-free growth. The ``six degrees of separation'' phenomenon \citep{Milgram1967, Watts1999} can be understood through our optimal detection regime, where super-linear growth creates bridging connections between clusters, developing short path lengths as nodes seek to maximise their information processing capabilities.

This super-linear growth in the optimal detection regime finds additional support in studies of scientific collaboration networks \cite{newman2003structure} and urban scaling relationships \cite{bettencourt2013origins}. These systems exhibit periods of accelerated growth in connectivity that exceed linear preferential attachment predictions before saturating. Recent studies of online social networks \cite{burke2016relationship} further show that user engagement and relationship quality follow patterns consistent with our predicted optimal detection regime, before declining in ways that suggest information processing saturation.

Our framework transforms qualitative insights into precise, testable predictions through the mathematical rigour of the Free Energy Principle. While existing empirical findings provide strong support, controlled experimental tests remain an important direction for future research. Particularly valuable would be experiments in artificial networks where information processing constraints could be varied while observing the resulting network formation patterns.

\subsection{Implications for Social Networks}

Our results have implications for the design of social networks, particularly for online platforms where users routinely operate far beyond natural information processing limits. The framework suggests that many of the concerns associated with social media \citep{Twenge2020, Przybylski2017} may stem from forcing users to function in the saturation regime (\(d \gg d_\star\)), where their cognitive systems cannot process the volume of social signals they receive. Modern platforms, with their continuous feeds, frequent notifications, and follower counts in the thousands or millions, push users well beyond the ``sweet spot'' where additional connections improve belief precision (\(\alpha d \gg \eta\) but \(d \ll d_\star\)). This aligns with recent findings showing decreased quality of social relationships and increased anxiety with excessive social media use \citep{Hunt2018}. In this saturated state, more social connections no longer enhance certainty about the social environment, potentially explaining documented decreases in the quality of online social interactions and decision-making \citep{Bakshy2015}.

These insights suggest several concrete interventions that could improve social network design. Platforms could implement strict limits on the maximum number of connections (\(k_{\text{max}}\)) that users can maintain, preventing them from entering the saturation regime. While perhaps controversial, such limits would align with known cognitive constraints like Dunbar's number \citep{Dunbar2016} and recent work on attention economics \citep{Wu2016}. More subtle approaches might include ``soft barriers'' that keep users operating in the optimal detection regime, similar to successful interventions in digital well-being \citep{Lukoff2018}. These could include capping daily interactions, limiting feed sizes to processable chunks, or creating natural breaking points in content consumption—strategies supported by research in cognitive load theory \citep{Sweller2019} and information overload \citep{Jones2004}.

The goal of such interventions would be to maintain users in the super-linear growth regime, where information processing is most effective. In this optimal zone, social connections provide genuine improvements in belief precision and action selection, possibly leading to more meaningful and satisfying online interactions \citep{Burke2016}. This perspective suggests that many apparent trade-offs between engagement and user well-being may be false dichotomies – by aligning platform design with basic principles of human information processing, we might create social networks that are both more engaging and more psychologically sustainable \citep{Meier2018}. These insights suggest how our theoretical framework can translate into practical recommendations for improving the design of social systems, bridging the gap between principles of biological information processing and real-world applications in digital social infrastructure \citep{Pennycook2019}.

Our framework generates testable predictions for social network design. For instance, we predict that in online environments characterized by higher sensory noise (e.g., overwhelming information flow, algorithmic filtering), social networks will exhibit fewer isolated nodes than predicted by pure preferential attachment, as users actively seek connections to reduce uncertainty. Conversely, in environments designed to limit individual information processing capacity (e.g., strict connection limits, attention management tools), we expect to observe fewer high-degree hubs and a more pronounced saturation effect in the degree distribution.

\subsection{Towards Networks of AI Agents}

Our theoretical framework offers insights for the emerging field of artificial agent networks, particularly as AI systems transition from isolated reasoning engines to interconnected, cooperative agents \citep{Dafoe2020}. The identification of distinct information processing regimes suggests important design considerations for multi-agent AI systems. Just as biological agents face constraints in processing social information, artificial agents may encounter analogous limitations in their ability to process and act on information from their peers. Understanding these constraints could be essential for creating stable and effective agent networks that avoid the pitfalls observed in human social networks.

The super-linear growth regime identified in our model has particular relevance for collaborative AI systems. Recent work in multi-agent reinforcement learning has shown that agent performance often scales non-linearly with the number of collaborative partners \citep{Baker2020}, but saturates or degrades with too many connections. Our framework suggests this pattern may reflect information processing constraints rather than mere implementation details. Just as agents have an optimal detection regime where $\alpha d \gg \eta$ but $d \ll d_\star$, artificial agents likely have an optimal zone of collaboration where they can process and act on information from their peers. This insight suggests that constraining agent networks to operate within this regime might improve collective performance and stability.

These principles could inform the design choices and architecture of large-scale AI systems \citep{Battaglia2018}. Rather than allowing unrestricted connections between agents, system designers might implement hierarchical structures or connection limits that keep individual agents operating in their optimal detection regime. Such designs would mirror successful biological systems, where information processing constraints have shaped network topology through evolutionary processes \citep{Mitchell2009}. Our framework suggests that the emergence of hub-and-spoke architectures in artificial neural systems \citep{LeCun2015} might reflect not just computational efficiency but principles of information processing under uncertainty.

As we move toward more complex systems of interacting AI agents \citep{Shoham2008}, understanding these network formation dynamics becomes ever more important. The principles identified here—-- particularly the balance between information seeking in sparse regimes, optimal processing in intermediate regimes, and saturation in dense regimes—-- could help prevent the emergence of pathological network structures that might lead to cascading failures or undesirable emergent behaviours \citep{Russell2019}. By considering information processing constraints in the design of multi-agent systems, we might create more robust and controllable artificial networks that maintain their performance even as they scale to larger sizes.

In the context of multi-agent AI systems, our model suggests that manipulating parameters related to information processing capacity, such as limiting the number of detectable peers or introducing artificial ``belief constraints'' in AI agents, could be used to engineer desired network structures. For example, reducing the information processing capacity of AI agents in collaborative tasks might, counterintuitively, lead to more scale-free network structures by mitigating saturation effects, allowing for the emergence of larger hubs. Conversely, increasing sensory noise in the agent's environment is predicted to drive agents to form denser, less hub-centric networks as they seek information more broadly.

\section{Conclusions}

The analysis presented here suggests a connection between the Free Energy Principle and the emergence of network structures, through a process that both explains and transcends traditional preferential attachment. We have shown that when agents act to minimise variational free energy, they can generate network structures with characteristics which can be attributed to preferential attachment; but with systematic and meaningful deviations that arise from information processing constraints. This result bridges two separate theoretical frameworks: the cognitive/biological framework of FEP \citep{Friston2010} and the network science framework of preferential attachment \citep{BarabasiAlbert1999}.

Our key theoretical results suggest that free energy minimising agents exhibit three distinct regimes of network formation. In the noise-dominated regime, where detection statistics are weak compared to noise $\alpha d \lesssim \eta$, agents seek better information, leading to fewer isolated nodes than pure preferential attachment would predict. In the optimal detection regime $\alpha d \gg \eta$ but $d \ll d_\star$, we have found a super-linear growth mechanism that emerges from the cascade of improving detection statistics, belief precision, and movement accuracy. This creates a characteristic concentration of nodes around optimal cluster sizes. Finally, in the saturation regime $d \gtrsim d_\star$, limitations in biological information processing prevent indefinite cluster growth, leading to fewer high-degree nodes than pure preferential attachment would predict. Together, these regimes explain the observed knee-shaped degree distributions in real networks as signatures of optimal information processing under biological constraints.

Many open questions remain. Future research could explore non-Gaussian noise models and richer observational processes, or consider how multiple temporal and spatial scales of inference interact to shape network growth. Extending the theory to multiplex or multilayer networks could also yield new insights, as could exploring how environmental non-stationarities alter the cognitive constraints and thus the resulting degree distributions. More broadly, investigating how additional external forces interact with FEP-driven behaviour would clarify when and how exogenous factors dominate over internal cognitive mechanisms.

The framework has relevant implications for understanding complex systems across scales. Our results suggest that deviations from pure scale-free structure in natural networks may be signatures of optimal information processing under constraints, rather than imperfections or anomalies. This perspective offers new approaches to analysing and designing collective systems, from biological networks to artificial swarms, by focusing on the information processing capabilities and limitations of their constituent agents.

The connection between FEP and network formation points to a deeper principle: complex systems may organise themselves to optimise information processing across scales. Wherever agents must process information to guide behaviour, the interplay between noise, optimal detection, and saturation regimes may shape the resulting network structures. This unifying perspective bridges individual cognition and collective structure, suggesting new directions for understanding emergence in complex systems.

Experimentally, our framework generates testable predictions about how information processing constraints shape network formation. Studies could measure the three key parameters ($k_{\text{max}}$, $b_{\text{max}}$, $v_{\text{max}}$) in real systems and examine how variations in these constraints affect resulting network structure. Particularly valuable would be experiments with artificial agent systems where information processing capabilities could be varied.

Future work should focus on testing these theoretical predictions in real systems and extending the framework to more complex scenarios. The relationship between cognitive parameters and network structure opens new avenues for empirical research, while the theoretical framework provides tools for designing self-organising systems with desired properties. Promising directions include applications to social media design, artificial agent networks, and biological collective behaviour.

In conclusion, this synthesis between FEP and network formation not only provides a mechanistic explanation for network structure but also suggests underlying principles governing the organisation of complex adaptive systems. By grounding network formation in the principles of uncertainty minimisation and information processing, we establish a theoretical foundation for understanding how cognitive constraints shape collective structure, with implications ranging from biological organisation to artificial system design.

\appendix
\section{Appendix: Closed Form Solutions for The Free Energy}
\label{appendix:closedform}

We show here how the free energy in Equation \ref{eq:free_energy_final} admits an explicit closed-form when all distributions are chosen to be Gaussian. Recall that our model is defined as
\begin{align}
\pi(b) &= \mathcal{N}\!\bigl(b; \mu_\pi, \sigma_\pi^2\bigr),\\
& \nonumber\\
L\bigl(\mathrm{Obs}\mid b\bigr) &= \mathcal{N}\!\bigl(D; \mu_D(b), \sigma_D^2\bigr), \\
& \nonumber\\
\mu_D(b) &= \alpha\,b + \beta,\\
& \nonumber\\
Q(b) &= \mathcal{N}\!\bigl(b; \mu_b, \sigma_b^2\bigr).
\end{align}
The free energy
\begin{align}
  F\bigl[Q(b)\bigr]
  \;=&\;
  \int Q(b)\,\ln\!\frac{Q(b)}{\pi(b)}\,db \nonumber\\
  \;&-\;
  \int Q(b)\,\ln L\!\bigl(\mathrm{Obs}\mid b\bigr)\,db,
\end{align}
splits into two integrals that can be computed separately.

The first integral is the Kullback--Leibler (KL) divergence between two Gaussians which has a known form. For \(\mathcal{N}(b;\,\mu_b,\sigma_b^2)\) and \(\mathcal{N}(b;\,\mu_\pi,\sigma_\pi^2)\) this is
\begin{equation}
\label{eq:KLgaussians}
  \mathrm{KL}\bigl[Q\|\pi\bigr] = \frac{1}{2} \Bigl[
    \ln \frac{\sigma_\pi^2}{\sigma_b^2} + \frac{\sigma_b^2 + (\mu_b - \mu_\pi)^2}{\sigma_\pi^2} - 1
  \Bigr].
\end{equation}
The second integral is the expected log-likelihood.
\begin{equation}
  -\mathbb{E}_Q\bigl[\ln L\bigr] = -\int Q(b)\ln \mathcal{N}\bigl(D;\mu_D(b),\sigma_D^2\bigr)\,db
\end{equation}
Because the natural log term in the integral has a quadratic dependence on \(\mu_D(b) = \alpha b + \beta\), we can write
\begin{equation}
  -\ln \mathcal{N}\bigl(D;\mu_D(b),\sigma_D^2\bigr) = \frac{\bigl(D - \mu_D(b)\bigr)^2}{2\sigma_D^2} +\text{const.}
\end{equation}
When \(Q(b)=\mathcal{N}(b;\,\mu_b,\sigma_b^2)\), this can be computed with a standard moment calculation.

Thus the combined expression for \(F\) is
\begin{align}
\label{eq:F_closedform}
  F\bigl[Q(b)\bigr] &=
  \underbrace{
    \frac{1}{2} \Bigl[ \ln \frac{\sigma_\pi^2}{\sigma_b^2} + \frac{\sigma_b^2 + (\mu_b - \mu_\pi)^2}{\sigma_\pi^2} - 1 \Bigr]
  }_{\mathrm{KL}[\,Q\|\pi\,]} \nonumber\\
  & \nonumber\\
  &+ \underbrace{
    \frac{ (D-\beta)^2 - 2\alpha (D-\beta) \mu_b + \alpha^2 \bigl(\mu_b^2+\sigma_b^2\bigr) }{2\sigma_D^2}  
  }_{-\mathbb{E}_Q\bigl[\ln L\bigr]} \nonumber\\
  &+\text{const.}
\end{align}
This is the full variational free energy under Gaussian prior, likelihood, and posterior. The notation ``\(\text{const.}\)'' hides terms that do not depend on \(\mu_b\). Minimising \(F\) with respect to $\mu_b$ yields update rules that balance prior alignment against fitting the observed detection statistic \(D\), under the assumption that $\sigma_b$ is held constant.

To see how the agent updates its posterior mean \(\mu_b\), we focus on the partial derivative \(\partial F/\partial \mu_b\). We start from the closed-form expression for \(F[Q(b)\bigr]\), Equation \ref{eq:F_closedform}, but note that \(\sigma_b^2\) must also be treated as a parameter to be optimised. For simplicity here, we only derive \(\tfrac{\partial F}{\partial \mu_b}\) and keep \(\sigma_b^2\) fixed; the joint update for \(\sigma_b^2\) could be derived similarly if required.

Differentiating Equation \ref{eq:F_closedform} with respect to $\mu_B$ gives
\begin{equation}
  \frac{\partial F}{\partial \mu_b} = \frac{\mu_b - \mu_\pi}{\sigma_\pi^2}
  +
  \frac{-\,\alpha\,(D-\beta) + \alpha^2\,\mu_b}{\sigma_D^2}.
\end{equation}
Setting \(\partial F/\partial \mu_b=0\) and solving for \(\mu_b\) yields the posterior mean that locally minimises \(F\). That closed-form solution is
\begin{equation}
\mu_b = \frac{(\alpha/\sigma_D^2)\,(D-\beta) \;+\; \mu_\pi/\sigma_\pi^2}{\alpha^2/\sigma_D^2 + 1/\sigma_\pi^2}.
\end{equation}
if $\sigma_b^2$ is held fixed.

\begin{acks}
This research was carried out at Rinna K.K., Tokyo, Japan.
\end{acks}
\bibliographystyle{SageH}
\bibliography{paper6}

\begin{thebibliography}{72}
\providecommand{\natexlab}[1]{#1}
\providecommand{\url}[1]{\texttt{#1}}
\providecommand{\urlprefix}{URL }
\expandafter\ifx\csname urlstyle\endcsname\relax
  \providecommand{\doi}[1]{DOI:\discretionary{}{}{}#1}\else
  \providecommand{\doi}{DOI:\discretionary{}{}{}\begingroup
  \urlstyle{rm}\Url}\fi

\bibitem[{Amaral et~al.(2000)Amaral, Scala, Barthélémy and
  Stanley}]{Amaral2000}
Amaral LAN, Scala A, Barthélémy M and Stanley HE (2000) Classes of
  small-world networks.
\newblock \emph{Proceedings of the National Academy of Sciences} 97(21):
  11149--11152.

\bibitem[{Anderson(2018)}]{Anderson2018}
Anderson PW (2018) More is different.
\newblock \emph{Science} 177: 393--396.

\bibitem[{Avena-Koenigsberger et~al.(2018)Avena-Koenigsberger, Mi{\v{s}}i{\'c}
  and Sporns}]{avena2018communication}
Avena-Koenigsberger A, Mi{\v{s}}i{\'c} B and Sporns O (2018) Communication
  dynamics in complex brain networks.
\newblock \emph{Nature Reviews Neuroscience} 19(1): 17--33.

\bibitem[{Axtell(2001)}]{Axtell2001}
Axtell RL (2001) Zipf distribution of u.s. firm sizes.
\newblock \emph{Science} 293(5536): 1818--1820.

\bibitem[{Baker et~al.(2020)Baker, Kanitscheider, Markov, Wu, Powell, McGrew
  and Mordatch}]{Baker2020}
Baker B, Kanitscheider I, Markov T, Wu Y, Powell G, McGrew B and Mordatch I
  (2020) Emergent tool use from multi-agent autocurricula.
\newblock \emph{International Conference on Learning Representations} .

\bibitem[{Bakshy et~al.(2015)Bakshy, Messing and Adamic}]{Bakshy2015}
Bakshy E, Messing S and Adamic LA (2015) Exposure to ideologically diverse news
  and opinion on facebook.
\newblock \emph{Science} 348(6239): 1130--1132.

\bibitem[{Barabási(2016)}]{Barabasi2016}
Barabási AL (2016) \emph{Network Science}.
\newblock Cambridge: Cambridge University Press.

\bibitem[{Barabási and Albert(1999)}]{BarabasiAlbert1999}
Barabási AL and Albert R (1999) Emergence of scaling in random networks.
\newblock \emph{Science} 286: 509--512.

\bibitem[{Bardoscia et~al.(2021)Bardoscia, Caccioli, Perotti, Vivaldo and
  Caldarelli}]{bardoscia2021physics}
Bardoscia M, Caccioli F, Perotti JI, Vivaldo G and Caldarelli G (2021) The
  physics of financial networks.
\newblock \emph{Nature Reviews Physics} 3(7): 490--507.

\bibitem[{Bascompte(2009)}]{Bascompte2009}
Bascompte J (2009) Disentangling the web of life.
\newblock \emph{Science} 325(5939): 416--419.

\bibitem[{Battaglia et~al.(2018)}]{Battaglia2018}
Battaglia PW et~al. (2018) Relational inductive biases, deep learning, and
  graph networks.
\newblock \emph{arXiv preprint arXiv:1806.01261} .

\bibitem[{Batty(2013)}]{Batty2013}
Batty M (2013) \emph{The New Science of Cities}.
\newblock MIT Press.

\bibitem[{Bettencourt(2013{\natexlab{a}})}]{Bettencourt2013}
Bettencourt LMA (2013{\natexlab{a}}) The origins of scaling in cities.
\newblock \emph{Science} 340(6139): 1438--1441.

\bibitem[{Bettencourt(2013{\natexlab{b}})}]{bettencourt2013origins}
Bettencourt LMA (2013{\natexlab{b}}) The origins of scaling in cities.
\newblock \emph{Science} 340(6139): 1438--1441.

\bibitem[{Bianconi and Barabási(2001)}]{Bianconi2001}
Bianconi G and Barabási AL (2001) Competition and multiscaling in evolving
  networks.
\newblock \emph{Europhysics Letters} 54: 436--442.

\bibitem[{Broido and Clauset(2019)}]{Broido2019}
Broido AD and Clauset A (2019) Scale-free networks are rare.
\newblock \emph{Nature Communications} 10: 1017.

\bibitem[{Burke and Kraut(2016{\natexlab{a}})}]{burke2016relationship}
Burke M and Kraut RE (2016{\natexlab{a}}) The relationship between facebook use
  and well-being depends on communication type and tie strength.
\newblock \emph{Journal of Computer-Mediated Communication} 21(4): 265--281.

\bibitem[{Burke and Kraut(2016{\natexlab{b}})}]{Burke2016}
Burke M and Kraut RE (2016{\natexlab{b}}) The relationship between facebook use
  and well-being depends on communication type and tie strength.
\newblock \emph{Journal of Computer-Mediated Communication} 21(4): 265--281.

\bibitem[{Clauset et~al.(2009)Clauset, Shalizi and Newman}]{Clauset2009}
Clauset A, Shalizi CR and Newman MEJ (2009) Power-law distributions in
  empirical data.
\newblock \emph{SIAM Review} 51: 661--703.

\bibitem[{Coase(1937)}]{Coase1937}
Coase RH (1937) The nature of the firm.
\newblock \emph{Economica} 4: 386--405.

\bibitem[{Colombo and Wright(2018)}]{Colombo2018}
Colombo M and Wright C (2018) First principles in the life sciences: the
  free-energy principle, organicism, and mechanism.
\newblock \emph{Synthese} 196: 5245--5267.

\bibitem[{Constant et~al.(2018)Constant, Ramstead, Veissière, Campbell and
  Friston}]{Constant2018}
Constant A, Ramstead MJD, Veissière SPL, Campbell JO and Friston KJ (2018) A
  variational approach to niche construction.
\newblock \emph{Journal of the Royal Society Interface} 15: 20170685.

\bibitem[{Couzin(2009)}]{Couzin2009}
Couzin ID (2009) Collective cognition in animal groups.
\newblock \emph{Trends in Cognitive Sciences} 13(2): 36--43.

\bibitem[{Dafoe et~al.(2020)Dafoe, Hughes, Bachrach, Collins, McKee, Leibo,
  Larson and Graepel}]{Dafoe2020}
Dafoe A, Hughes E, Bachrach Y, Collins T, McKee KR, Leibo JZ, Larson K and
  Graepel T (2020) Open problems in cooperative {AI}.
\newblock \emph{arXiv preprint arXiv:2012.08630} .

\bibitem[{Dorogovtsev and Mendes(2002)}]{Dorogovtsev2002}
Dorogovtsev SN and Mendes JFF (2002) Evolution of networks.
\newblock \emph{Advances in Physics} 51(4): 1079--1187.

\bibitem[{Dunbar(2016{\natexlab{a}})}]{dunbar2016human}
Dunbar R (2016{\natexlab{a}}) \emph{Human Evolution: Our Brains and Behavior}.
\newblock Oxford University Press.

\bibitem[{Dunbar(2016{\natexlab{b}})}]{Dunbar2016}
Dunbar R (2016{\natexlab{b}}) \emph{Human Evolution: Our Brains and Behavior}.
\newblock Oxford University Press.

\bibitem[{Dunbar(1992)}]{Dunbar1992}
Dunbar RIM (1992) Neocortex size as a constraint on group size in primates.
\newblock \emph{Journal of Human Evolution} 22(6): 469--493.

\bibitem[{England(2015)}]{England2015}
England JL (2015) Dissipative adaptation in driven self-assembly.
\newblock \emph{Nature Nanotechnology} 10: 919--923.

\bibitem[{Flack(2012)}]{Flack2012}
Flack JC (2012) Multiple time-scales and the developmental dynamics of social
  systems.
\newblock \emph{Philosophical Transactions of the Royal Society B} 367(1597):
  1802--1810.

\bibitem[{Friston(2010)}]{Friston2010}
Friston K (2010) The free-energy principle: a unified brain theory?
\newblock \emph{Nature Reviews Neuroscience} 11: 127--138.

\bibitem[{Friston(2013)}]{Friston2013}
Friston K (2013) Life as we know it.
\newblock \emph{Journal of the Royal Society Interface} 10: 20130475.

\bibitem[{Friston et~al.(2017)Friston, FitzGerald, Rigoli, Schwartenbeck and
  Pezzulo}]{Friston2017}
Friston K, FitzGerald T, Rigoli F, Schwartenbeck P and Pezzulo G (2017) Active
  inference: A process theory.
\newblock \emph{Neural Computation} 29: 1--49.

\bibitem[{Gollo et~al.(2018)Gollo, Roberts and Cocchi}]{gollo2018diversity}
Gollo LL, Roberts JA and Cocchi L (2018) Mapping how local perturbations
  influence systems-level brain dynamics.
\newblock \emph{NeuroImage} 160: 97--112.

\bibitem[{Hamann(2018)}]{Hamann2018}
Hamann H (2018) Swarm robotics: A formal approach.
\newblock \emph{Springer International Publishing} .

\bibitem[{Hunt et~al.(2018)Hunt, Marx, Lipson and Young}]{Hunt2018}
Hunt MG, Marx R, Lipson C and Young J (2018) No more fomo: Limiting social
  media decreases loneliness and depression.
\newblock \emph{Journal of Social and Clinical Psychology} 37(10): 751--768.

\bibitem[{Jackson(2010)}]{Jackson2010}
Jackson MO (2010) \emph{Social and Economic Networks}.
\newblock Princeton University Press.

\bibitem[{Jelasity and Babaoglu(2007)}]{Jelasity2007}
Jelasity M and Babaoglu O (2007) T-man: Gossip-based overlay topology
  management.
\newblock \emph{Engineering Self-Organising Systems} 3910: 1--15.

\bibitem[{Jeong et~al.(2003{\natexlab{a}})Jeong, Mason, Barab{\'a}si and
  Oltvai}]{jeong2003lethality}
Jeong H, Mason S, Barab{\'a}si AL and Oltvai Z (2003{\natexlab{a}}) Lethality
  and centrality in protein networks.
\newblock \emph{Nature} 411: 41--42.

\bibitem[{Jeong et~al.(2003{\natexlab{b}})Jeong, Mason, Barabási and
  Oltvai}]{Jeong2003}
Jeong H, Mason SP, Barabási AL and Oltvai ZN (2003{\natexlab{b}}) Lethality
  and centrality in protein networks.
\newblock \emph{Nature} 411: 41--42.

\bibitem[{Jones et~al.(2004)Jones, Ravid and Rafaeli}]{Jones2004}
Jones Q, Ravid G and Rafaeli S (2004) Information overload and the message
  dynamics of online interaction spaces.
\newblock \emph{Information Systems Research} 15(2): 194--210.

\bibitem[{Kirchhoff et~al.(2018)Kirchhoff, Parr, Palacios, Friston and
  Kiverstein}]{Kirchhoff2018}
Kirchhoff M, Parr T, Palacios E, Friston K and Kiverstein J (2018) The markov
  blankets of life: autonomy, active inference and the free energy principle.
\newblock \emph{Journal of the Royal Society Interface} 15: 20170792.

\bibitem[{Kossinets and Watts(2006)}]{kossinets2006empirical}
Kossinets G and Watts DJ (2006) Empirical analysis of an evolving social
  network.
\newblock \emph{Science} 311(5757): 88--90.

\bibitem[{LeCun et~al.(2015)LeCun, Bengio and Hinton}]{LeCun2015}
LeCun Y, Bengio Y and Hinton G (2015) Deep learning.
\newblock \emph{Nature} 521(7553): 436--444.

\bibitem[{Lukoff et~al.(2018)Lukoff, Yu, Kientz and Hiniker}]{Lukoff2018}
Lukoff K, Yu C, Kientz J and Hiniker A (2018) What makes smartphone use
  meaningful or meaningless?
\newblock \emph{Proceedings of the ACM on Interactive, Mobile, Wearable and
  Ubiquitous Technologies} 2(1): 1--26.

\bibitem[{Lynn and Bassett(2020{\natexlab{a}})}]{Lynn2020}
Lynn CW and Bassett DS (2020{\natexlab{a}}) The physics of brain network
  structure, function and control.
\newblock \emph{Nature Reviews Physics} 1: 318--332.

\bibitem[{Lynn and Bassett(2020{\natexlab{b}})}]{lynn2020physics}
Lynn CW and Bassett DS (2020{\natexlab{b}}) The physics of brain network
  structure, function and control.
\newblock \emph{Nature Physics} 16(9): 965--976.

\bibitem[{Medo et~al.(2011)Medo, Cimini and Gualdi}]{Medo2011}
Medo M, Cimini G and Gualdi S (2011) Temporal effects in the growth of
  networks.
\newblock \emph{Physical Review Letters} 107: 238701.

\bibitem[{Meier and Reinecke(2018)}]{Meier2018}
Meier A and Reinecke L (2018) Computer-mediated communication, social media,
  and mental health: A conceptual and empirical meta-review.
\newblock \emph{Communication Research} 45(8): 1182--1209.

\bibitem[{Milgram(1967)}]{Milgram1967}
Milgram S (1967) The small-world problem.
\newblock \emph{Psychology Today} 1(1): 61--67.

\bibitem[{Mitchell(2009)}]{Mitchell2009}
Mitchell M (2009) \emph{Complexity: A Guided Tour}.
\newblock Oxford University Press.

\bibitem[{Newman(2003{\natexlab{a}})}]{newman2003structure}
Newman ME (2003{\natexlab{a}}) The structure and function of complex networks.
\newblock \emph{SIAM Review} 45(2): 167--256.

\bibitem[{Newman(2003{\natexlab{b}})}]{Newman2003}
Newman MEJ (2003{\natexlab{b}}) The structure and function of complex networks.
\newblock \emph{SIAM Review} 45: 167--256.

\bibitem[{Newman(2005)}]{Newman2005}
Newman MEJ (2005) Power laws, pareto distributions and zipf's law.
\newblock \emph{Contemporary Physics} 46: 323--351.

\bibitem[{Newman(2018)}]{Newman2018}
Newman MEJ (2018) \emph{Networks}.
\newblock Oxford University Press.

\bibitem[{Olesen et~al.(2007)Olesen, Bascompte, Dupont and
  Jordano}]{Olesen2007}
Olesen JM, Bascompte J, Dupont YL and Jordano P (2007) The modularity of
  pollination networks.
\newblock \emph{Proceedings of the National Academy of Sciences} 104(50):
  19891--19896.

\bibitem[{Papadopoulos et~al.(2012{\natexlab{a}})Papadopoulos, Kitsak, Serrano,
  Bogu\~{n}\'{a} and Krioukov}]{Papadopoulos2012}
Papadopoulos F, Kitsak M, Serrano MA, Bogu\~{n}\'{a} M and Krioukov D
  (2012{\natexlab{a}}) Popularity versus similarity in growing networks.
\newblock \emph{Nature} 489: 537--540.

\bibitem[{Papadopoulos et~al.(2012{\natexlab{b}})Papadopoulos, Kitsak, Serrano,
  Bogu{\~n}{\'a} and Krioukov}]{papadopoulos2012popularity}
Papadopoulos F, Kitsak M, Serrano M{\'A}, Bogu{\~n}{\'a} M and Krioukov D
  (2012{\natexlab{b}}) Popularity versus similarity in growing networks.
\newblock \emph{Nature} 489(7417): 537--540.

\bibitem[{Parr et~al.(2020)Parr, Da~Costa and Friston}]{Parr2020}
Parr T, Da~Costa L and Friston K (2020) Markov blankets, information geometry
  and stochastic thermodynamics.
\newblock \emph{Philosophical Transactions of the Royal Society A} 378:
  20190159.

\bibitem[{Pennycook and Rand(2019)}]{Pennycook2019}
Pennycook G and Rand DG (2019) Fighting misinformation on social media using
  crowdsourced judgments of news source quality.
\newblock \emph{Proceedings of the National Academy of Sciences} 116(7):
  2521--2526.

\bibitem[{Przybylski and Weinstein(2017)}]{Przybylski2017}
Przybylski AK and Weinstein N (2017) A large-scale test of the goldilocks
  hypothesis: Quantifying the relations between digital-screen use and the
  mental well-being of adolescents.
\newblock \emph{Psychological Science} 28(2): 204--215.

\bibitem[{Ramstead et~al.(2018)Ramstead, Badcock and Friston}]{Ramstead2018}
Ramstead MJD, Badcock PB and Friston KJ (2018) Answering schrödinger's
  question: A free-energy formulation.
\newblock \emph{Physics of Life Reviews} 24: 1--16.

\bibitem[{Russell(2019)}]{Russell2019}
Russell S (2019) \emph{Human Compatible: Artificial Intelligence and the
  Problem of Control}.
\newblock Viking.

\bibitem[{Schweitzer et~al.(2009)Schweitzer, Fagiolo, Sornette, Vega-Redondo,
  Vespignani and White}]{Schweitzer2009}
Schweitzer F, Fagiolo G, Sornette D, Vega-Redondo F, Vespignani A and White DR
  (2009) Economic networks: The new challenges.
\newblock \emph{Science} 325(5939): 422--425.

\bibitem[{Shoham and Leyton-Brown(2008)}]{Shoham2008}
Shoham Y and Leyton-Brown K (2008) Multiagent systems: Algorithmic,
  game-theoretic, and logical foundations.
\newblock \emph{Cambridge University Press} .

\bibitem[{Sumpter(2010{\natexlab{a}})}]{Sumpter2010}
Sumpter DJT (2010{\natexlab{a}}) Collective animal behavior.
\newblock \emph{Princeton University Press} .

\bibitem[{Sumpter(2010{\natexlab{b}})}]{sumpter2010collective}
Sumpter DJT (2010{\natexlab{b}}) Collective animal behavior.
\newblock \emph{Princeton University Press} .

\bibitem[{Sweller(2019)}]{Sweller2019}
Sweller J (2019) Cognitive load theory and educational technology.
\newblock \emph{Educational Technology Research and Development} 67: 1--16.

\bibitem[{Twenge(2020)}]{Twenge2020}
Twenge JM (2020) Why increases in adolescent depression may be linked to the
  technological environment.
\newblock \emph{Current Opinion in Psychology} 32: 89--94.

\bibitem[{Ver~Steeg(2019)}]{Ver2019}
Ver~Steeg G (2019) Entropy and information in neural networks.
\newblock \emph{Physical Review Letters} 123: 178301.

\bibitem[{Watts(1999)}]{Watts1999}
Watts DJ (1999) \emph{Small Worlds: The Dynamics of Networks between Order and
  Randomness}.
\newblock Princeton University Press.

\bibitem[{Wu(2016)}]{Wu2016}
Wu T (2016) \emph{The Attention Merchants: The Epic Scramble to Get Inside Our
  Heads}.
\newblock Knopf.

\end{thebibliography}
\end{document}